\documentclass[aps,prb,twocolumn,floatfix,amsmath,amssymb]{revtex4}
\usepackage{graphicx}
\usepackage{color}
\usepackage{dcolumn}
\usepackage{bm}
\usepackage{epsfig,psfrag,amsmath,amssymb,float}
\input{epsf}

\begin{document}

\title{Strong enhancement of the d-wave superconducting state
in the three-band Hubbard model coupled to an apical oxygen phonon}
\author{Z.~B.~Huang$^1$, H.~Q.~Lin$^2$, and E.~Arrigoni$^3$}
\affiliation{$^1$Faculty of Physics and Electronic Technology, Hubei
University, Wuhan 430062, China}
\affiliation{$^2$Department of Physics, the Chinese University of
Hong Kong, Hong Kong, China}
\affiliation{$^3$Institute of Theoretical and Computational Physics,
Graz University of Technology, Petersgasse 16, A-8010 Graz, Austria}
\date{\today}

\begin{abstract}
We study the hole binding energy and pairing correlations in the
three-band Hubbard model coupled to an apical oxygen phonon, by
exact diagonalization and constrained-path Monte Carlo simulations.
In the physically relevant charge-transfer regime, we find that the
hole binding energy is strongly enhanced by the electron-phonon
interaction, which is due to a novel potential-energy-driven pairing
mechanism involving reduction of both electronic potential energy
and phonon related energy. The enhancement of hole binding energy,
in combination with a phonon-induced increase of quasiparticle
weight, leads to a dramatic enhancement of the long-range part of
d-wave pairing correlations. Our results indicate that the apical
oxygen phonon plays a significant role in the superconductivity of
high-$T_c$ cuprates.
\end{abstract}

\pacs{PACS Numbers: 74.20.-z, 71.10.Fd, 74.25.Kc, 74.72.-h}
\maketitle
\section{Introduction}
Despite years of intensive research, the pairing
mechanism responsible for d-wave superconductivity (dSC) in the
high-$T_c$ cuprates remains a puzzle. It is generally believed that
the conventional phonon mechanism is inconsistent with d-wave
pairing symmetry and not strong enough to explain transition
temperatures higher than 100~K. Most investigations in this
direction have been focused on the pure electronic mechanism, but no
consensus has been reached so far. Recently, accurate experiments
displayed pronounced phonon and electron-lattice effects in these
materials, which are manifested by a large softening and broadening
of certain phonon modes in the whole doping region. In particular,
the in-plane copper-oxygen bond-stretching phonon, apical oxygen
phonon (AOP), and oxygen $B_{1g}$ buckling phonon are shown to be
strongly coupled to charge carriers \cite{pint1,pint2}. Moreover,
photoemission-spectroscopy-resolved kink structures
~\cite{lanzara,cuk} are probably caused by coupling of
quasiparticles to phonon modes. These findings suggest that phonons
are important for the physical properties of high-$T_c$ cuprates.

Various theoretical attempts have been made to understand the role
of phonons in high-$T_c$ superconductivity
(HTSC)~\cite{huang1B,castern,vidmar,jarrell}, but the answer remains
unclear. In a functional renormalization group study~\cite{castern},
Honerkamp {\it et al.} found that the $B_{1g}$ buckling phonon
enhances the d-wave pairing instability in the Hubbard model. More
recently, an exact diagonalization (ED) study of the $t-J$ model
coupled to phonons shows that coupling to the buckling mode
stabilizes d-wave pairing while coupling to the breathing mode
favors a p-wave pairing~\cite{vidmar}. On the other hand, based on
dynamical cluster Monte Carlo calculations of the Hubbard model
coupled to Holstein, buckling and breathing phonons, Macridin {\it
et al.} found that while these phonons can indeed enhance pairing, a
strong phonon-induced reduction of quasiparticle weight leads to a
suppression of dSC~\cite{jarrell}.

In this paper, we study the effect of AOP on dSC in the more
realistic three-band Hubbard model. Our work is motivated by recent
experiments showing that the distance between apical oxygen and the
$CuO_2$ plane~\cite{slezak}, the disorder around apical
oxygen~\cite{hobou,gao}, and the apical hole state~\cite{merz} have
significant effects on $T_c$. Basically, there are two routes for
apical oxygen to affect HTSC: one is to tune the electronic
structure of $CuO_2$ plane, leading to a change of
$T_c$~\cite{mori}; the other one is to directly couple apical oxygen
vibrations to charge carriers on the conducting $CuO_2$
plane~\cite{frick3B,frick1B}. Here, we focus on the second route and
study a strongly anharmonic vibration of apical oxygen in a
double-well potential, which is evidenced in the X-ray absorption
spectroscopy of several typical high-$T_c$
compounds~\cite{haskel,steven,leon}.
For a strongly anharmonic motion
in the double-well potential, the first excitation energy
$\Delta E=E_{1}-E_{0}$ is much smaller than the ones excited to
higher energy levels $E_{n \ge 2}$. In this case one can take into
account only the lowest quantum states $\Phi_{0}$ and $\Phi_{1}$
with energies $E_{0}$ and $E_{1}$ and model the low-energy motion
of apical oxygen by a local two-level system represented by a
pseudospin~\cite{plakida1,plakida2} degree of freedom.

Our main results, obtained by ED and constrained-path Monte Carlo
(CPMC) methods, are presented in Figs.~\ref{sum2x2}(b),
\ref{bindE}(a) and \ref{PdVdU246}. Fig.~\ref{bindE}(a) clearly shows
that the coupling to the AOP induces a strong enhancement of hole
binding energy, and this enhancement effect grows as the Coulomb
repulsion $U_d$ on the copper site is increased. An analysis of the
contribution of different energies to the hole binding energy
reveals a novel potential-energy-driven pairing mechanism that
involves reduction of both electronic potential energy and phonon
related energy. As a combination of increasing pairing interaction
and quasiparticle weight (see Fig.~\ref{sum2x2}(b)), the d-wave
pairing correlations are found to be strongly enhanced by the
electron-phonon (el-ph) coupling (see Fig.~\ref{PdVdU246}).

Our paper is organized as follows: In Section~\ref{Model},
we define the Hamiltonian and the physical quantities
calculated and discuss the choice of model parameters.
In Section~\ref{Results}, we present our numerical results and discuss
the physical mechanism responsible for the AOP-induced enhancement of
dSC. Finally, in Section~\ref{Conclusions}, we discuss in detail
our main conclusions.

\section{\label{Model} Model and numerical approach}
To model the electronic structure of $CuO_2$ plane and the coupling
of holes to the anharmonic AOP, we adopt the following Hamiltonian
proposed in Ref.~\cite{frick3B},
\begin{eqnarray}
H&=&H_{k}+H_{pot}+H_{ph},
\end{eqnarray}
where $H_{k}$, $H_{pot}$, and $H_{ph}$ stand for the kinetic
motion of holes, the potential energy for holes, and phonon related
energy, respectively. They are expressed in the form:
\begin{eqnarray}
H_{k}&=&\sum_{\langle i,j\rangle\sigma}t_{pd}^{ij}
(d_{i\sigma}^{\dagger}p_{j\sigma}+h.c.)\nonumber\\
&&+\sum_{\langle j,k\rangle\sigma}
t_{pp}^{jk}(p_{j\sigma}^{\dagger}
p_{k\sigma}+h.c.),
\end{eqnarray}
\begin{eqnarray}
H_{pot}&=&\epsilon\sum_{j\sigma}n_{j\sigma}^{p}
+U_{d}\sum_{i}n_{i\uparrow}^{d}n_{i\downarrow}^{d},
\end{eqnarray}
and
\begin{eqnarray}
\label{Hph}
H_{ph}&=&\sum_{i}(g_{cu}n_{i}^{d}+g_{o}
\sum_{\delta'}n_{i+\delta'}^{p})s_{i}^{z}
-\Omega\sum_{i}s_{i}^{x},
\end{eqnarray}
Here, the operator $d_{i\sigma}^{\dagger}$ creates a hole at a Cu
$3d_{x^2-y^2}$ orbital and $p_{j\sigma}^{\dagger}$ creates a hole in
an O$2p_x$ or $2p_y$ orbital. $s_{i}^{z}$ and $s_{i}^{x}$ are
the pseudospin operators for $S=1/2$.
$U_d$ denotes the Coulomb energy at
the Cu sites. $t_{pd}^{ij} = \pm t_{pd}$ and $t_{pp}^{jk} = \pm
t_{pp}$ are the Cu-O and O-O hybridizations, respectively, with the
Cu and O orbital phase factors included in the sign. The
charge-transfer energy is $\epsilon$, i.e. the oxygen orbital
energy. $g_{cu}$ and $g_{o}$ in $H_{ph}$ denote the strength of
el-ph coupling. $\Omega$ stands for the tunneling frequency of the
two-level system. $\delta'$ is the vector connecting Cu and its
nearest-neighbor (NN) O. According to quantum cluster
calculations~\cite{martin}, the parameters lie in the range:
$t_{pd}=1.3-1.5eV$, $t_{pp}=0.6eV$, $\epsilon=3.6eV$, and
$U_{d}=8.5-10.5eV$. In units of $t_{pd}$, we choose a parameter set
$t_{pp}=0.3$ and $\epsilon=3$, while $U_d$ is varied from weak to
strong coupling, including the physical value $U_{d}=6-8$.
$\Omega=0.5$ and $g_{cu}=g_{o}=g$ are assumed for the results
presented below, except explicitly noted otherwise.

Our calculations are performed on clusters of $2 \times 2$,
$8 \times 4$, $6 \times 6$ and $8 \times 6$
unit cells with periodic boundary conditions using the ED and CPMC
methods~\cite{shiwei,huang3B}. In the CPMC method, we follow
Refs.~\onlinecite{frick1B} and \onlinecite{frick3B}
to use the Worldline representation for
the pseudospins and projected the ground state $|\Psi_0 \rangle$ of
el-ph interacting system from a trial wave function
$|\Psi_{T}\rangle=|\Psi_{e}^{T}\rangle \otimes
|\Psi_{s}^{T}\rangle$. Here, $|\Psi_{e}^{T}\rangle$ and
$|\Psi_{s}^{T}\rangle$ represent the hole and phonon parts,
respectively. The CPMC algorithm has been checked against ED on the
$2\times 2$ cluster, and the difference for the electronic kinetic
energy, as well as for the charge and magnetic moment at the copper
sites, is less than $3\%$ up to $U_{d}=6$.

The hole binding energy is defined as:
\begin{equation}
\label{binding}
\Delta = E_{2} + E_{0} - 2 E_{1},
\end{equation}
with $E_n $ the ground-state energy for n doped holes.
The $d_{x^2-y^2}$ pairing correlation
is defined by,
\begin{equation}
\label{pairing}
P_{d}(\vec R) = \langle\Delta_{d}^\dagger(\vec
R) \Delta_{d}(0)\rangle,
\end{equation}
where
\begin{eqnarray*}
\Delta_{d}(\vec R) = \sum\limits_{\vec{\delta}}
f_{d}(\vec{\delta}) \{&[&d_{\vec{R}\uparrow}d_{\vec{R}+\vec{\delta}
\downarrow} -d_{\vec{R}\downarrow}d_{\vec{R}+\vec{\delta}\uparrow}]\\
+ &[&p^x_{\vec{R}\uparrow}p^x_{\vec{R}+\vec{\delta}\downarrow}
-p^x_{\vec{R}\downarrow}p^x_{\vec{R}+\vec{\delta}\uparrow} ]\\
+&[&p^y_{\vec{R}\uparrow}p^y_{\vec{R}+\vec{\delta}\downarrow}
-p^y_{\vec{R}\downarrow}p^y_{\vec{R}+\vec{\delta}\uparrow}] \}
\end{eqnarray*}
with $\vec{\delta} = \pm \hat{x}, \pm \hat{y}$.
$f_{d}(\vec{\delta}) = 1 $ for $\vec{\delta} = \pm \hat{x}$
and $f_{d}(\vec{\delta}) = -1 $ for $\vec{\delta} = \pm \hat{y} $.
We calculate also the vertex
contribution to the correlations defined as follows:
\begin{equation}
V_{d}(\vec R) = P_{d}(\vec R) - \bar{P}_{d}(\vec R),
\end{equation}
where $\bar{P}_{d}(\vec R)$ is the bubble contribution obtained
with the dressed (interacting) propagator~\cite{white}.

\section{\label{Results} Results and Discussions}
\begin{center}
\begin{figure}
\epsfig{file=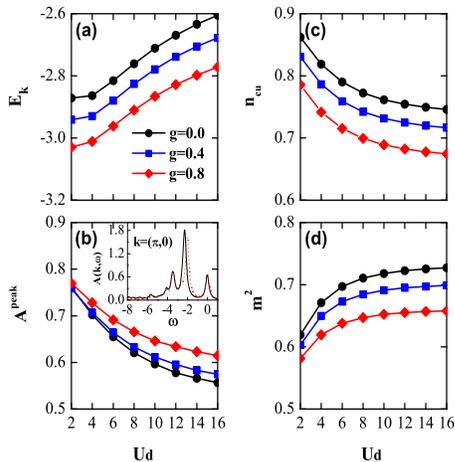,height=7.0cm,width=6.8cm,angle=0}
\medskip
\caption{(color online)(a) Electronic kinetic energy $E_{k}$ (per
unit cell); (b) peak value $A^{peak}$ of the spectral function
$A(k=(\pi,0),\omega)$ at the Fermi energy; (c) charge $n_{cu}$ and
(d) magnetic moment $m^2$ at the copper sites as a function of $U_d$
at different el-ph coupling $g$. The inset of figure (b) displays
$A(k=(\pi,0),\omega)$ as a function of $\omega$ at $U_{d}=6$, with
$g=0.0$ (solid line) and $g=0.8$ (dashed line). The The results are
obtained from ED for the $2 \times 2$ cluster.} \label{sum2x2}
\end{figure}
\end{center}

First, we show ED results for the $2 \times 2$ cluster with one hole
doped beyond half filling. The electronic kinetic energy
$E_{k}=\langle H_{k}\rangle$, the peak value $A^{peak}$ of the
single-particle spectral function $A(k,\omega)$ at the Fermi energy
($\omega=0$), the charge $n_{cu}=\langle n_{i}^{d}\rangle$ and the
magnetic moment
$m^2=\langle(n_{i\uparrow}^{d}-n_{i\downarrow}^{d})^2\rangle$ at the
copper sites are displayed in Figs.~\ref{sum2x2}(a)-\ref{sum2x2}(d).
Here, $A(k,\omega)=-\frac{1}{\pi}\sum_{m}
\langle\Phi_{0}(1)|c_{m,k\sigma}(\omega-E_{1}+H+i\eta)^{-1}
c_{m,k\sigma}^{\dagger}|\Phi_{0}(1)\rangle$ with $\Phi_{0}(1)$ and
$E_1$ denoting the ground-state wave function and its energy for
one-hole doping. The index $m$ corresponds either to the
$d_{x^2-y^2}$ or to the $p_{x,y}$ orbitals, and $\eta=0.2$. One can
clearly see that $E_{k}$ is lowered with increasing the el-ph
coupling $g$ at all Coulomb energies. Meanwhile, an increase of
$A^{peak}$ with increasing $g$ indicates that the quasiparticle
weight is increased by the el-ph coupling. In contrast, previous
studies of harmonic phonons in the Holstein-Hubbard model found that
the kinetic energy of electrons is increased with increasing el-ph
coupling, accompanying a reduction of quasiparticle
weight~\cite{jarrell,gunnarsson}. To explore the physical reasons
for lowering $E_{k}$, we switch off the coupling of apical phonon
either to copper or to in-plane oxygen, i.e., we set $g_{cu}=0$ or
$g_{o}=0$ in Eq.(\ref{Hph}). It is found that $E_{k}$ is lowered for
the former case, but increased for the latter case. These results
demonstrate that the special coupling of apical oxygen phonon to
in-plane oxygen is responsible for lowering the electronic kinetic
energy.

\begin{table}[b]
\caption{$g$ dependence of $E_{k}$ (per unit cell), $n_{cu}$, $m^2$,
and NN $Cu-Cu$ spin correlation $\langle S_{i}^{z}\cdot
S_{j}^{z}\rangle$ on the $6\times 6$ cluster at $U_{d}=6$. The
number of holes $Nh=42$, corresponding to a hole doping density $x
\sim 0.167$. Statistical errors are in the last digit and shown in
the parentheses.}
\begin{tabular}{p{1.2cm}p{1.6cm}p{1.6cm}p{1.6cm}p{1.6cm}}
\hline\hline
$g$ & $E_{k}$ & $n_{cu}$ & $m^2$ & $\langle
S_{i}^{z}\cdot S_{j}^{z}\rangle$\\
\hline
0.0 & -2.612(2) & 0.7563(1) & 0.6937(1) & -0.0800(3)\\
0.2 & -2.659(3) & 0.7426(2) & 0.6786(1) & -0.0715(3)\\
0.4 & -2.691(6) & 0.7292(3) & 0.6651(2) & -0.0659(6)\\
0.6 & -2.723(9) & 0.7142(5) & 0.6505(5) & -0.0605(9)\\
\hline\hline
\end{tabular}
\label{tab6x6}
\end{table}

From Fig.~\ref{sum2x2}(c), we notice that the charge is transferred
from copper to oxygen sites, which, in combination with a
phonon-mediated retarded attraction between holes with opposite
spins, results in a reduction of magnetic moment at $Cu$ sites, as
shown in Fig.~\ref{sum2x2}(d). Similar effects of AOP on $E_{k}$,
$n_{cu}$ and $m^2$ are also observed for larger clusters obtained by
CPMC simulations, and representative results on the $6 \times 6$
cluster are shown in Table~\ref{tab6x6}. The last column in
Tab.~\ref{tab6x6} shows that the value of NN $Cu-Cu$ spin
correlation $\langle S_{i}^{z}\cdot S_{j}^{z}\rangle$ becomes less
negative with increasing $g$, implying a suppression of
antiferromagnetic (AFM) spin correlation.

\begin{center}
\begin{figure}
\epsfig{file=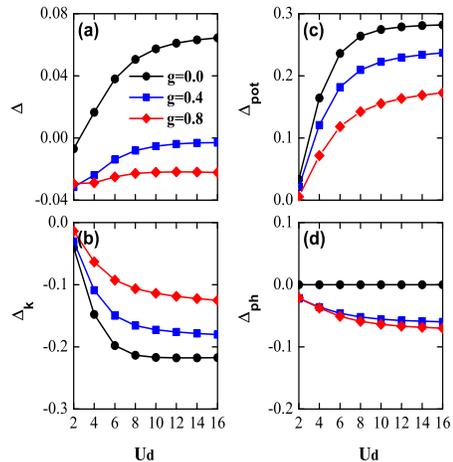,height=7cm,width=6.8cm,angle=0}
\caption{(color online) (a) Hole binding energy $\Delta$ obtained
from ED for the $2 \times 2$ cluster as a function of $U_d$. (b)-(d)
show the contributions to $\Delta$ from different energies, as
discussed in text. The el-ph coupling $g$ is indicated by the shape
of the symbol.} \label{bindE}
\end{figure}
\end{center}

The hole binding energy $\Delta$ is shown in Fig.~\ref{bindE}(a)
as a function of the Coulomb energy $U_d$ at different $g$.
At all $U_d$, the binding energy is decreased by switching on
the el-ph coupling, signaling an enhancement of hole pairing
interaction. It is remarkable that this enhancement effect
becomes stronger with increasing $U_d$,
which is particularly evident in the region $U_{d}\le 8$.

In order to identify the physical origin for this enhancement, the
contributions to $\Delta$ from the hole kinetic energy $E_{k}$, the
hole potential energy $E_{pot}=\langle H_{pot}\rangle$, and the
phonon related energy $E_{ph}=\langle H_{ph}\rangle$ are depicted in
Figs.~\ref{bindE} (b)-(d), respectively. Here, $\Delta_{k}$,
$\Delta_{pot}$ and $\Delta_{ph}$ have similar definitions to
$\Delta$, with $E$ in Eq.(\ref{binding}) replaced with $E_{k}$,
$E_{pot}$ and $E_{ph}$, respectively. These quantities represent the
gain in the corresponding energy when the second hole is doped in
the vicinity of the first one. Although the kinetic energy of holes
is reduced upon increasing the el-ph coupling (see
Fig.~\ref{sum2x2}(a)), an increase of $\Delta_{k}$ with increasing
$g$ displayed in Fig.~\ref{bindE}(b) indicates that the kinetic
energy gain for two doped holes is reduced by the el-ph coupling. As
seen in Fig.~\ref{bindE}(c) and Fig.~\ref{bindE}(d), a decrease of
$\Delta_{pot}$ and $\Delta_{ph}$ with increasing $g$ reveals that it
is the reduction of electronic potential energy and phonon related
energy between two doped holes that enhances the hole binding energy
$\Delta$. In addition, the decrease of $\Delta_{pot}$ and
$\Delta_{ph}$ becomes more pronounced as $U_d$ is increased, leading
to a stronger enhancement of the hole binding energy in the strong
correlation regime (see Fig.~\ref{bindE}(a)).

\begin{center}
\begin{figure}
\epsfig{file=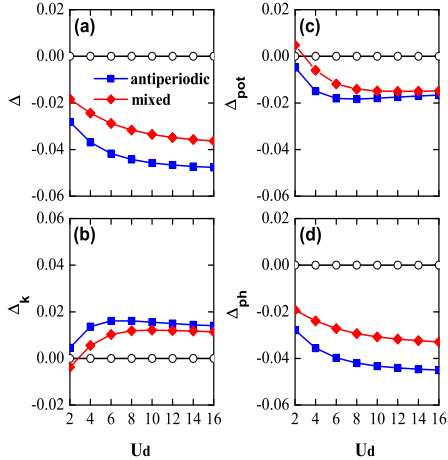,height=7cm,width=6.8cm,angle=0}
\caption{(color online) (a) Hole binding energy $\Delta$ obtained
from ED for the $2 \times 2$ cluster as a function of $U_d$. (b)-(d)
show the contributions to $\Delta$ from different energies.
Open circle is for $g=0$, and filled symbols for $g=0.4$ with
antiperiodic (square) and mixed (diamond) boundary conditions,
respectively.} \label{bindE2}
\end{figure}
\end{center}

Fig.~\ref{bindE2} displays the hole binding energy and different
contributions to $\Delta$ on the $2\times 2$ cluster with
antiperiodic and mixed (periodic in the x direction and antiperiodic
in the y direction) boundary conditions. A comparison of the results
in Figs.~\ref{bindE} and \ref{bindE2} shows that although the boundary
condition has strong effects on the amplitude of hole binding energy,
the AOP-induced enhancement of hole binding energy is qualitatively
similar for different boundary conditions. This demonstrates that
our findings reflect the intrinsic effects of AOP in the studied model.

\begin{center}
\begin{figure}
\epsfig{file=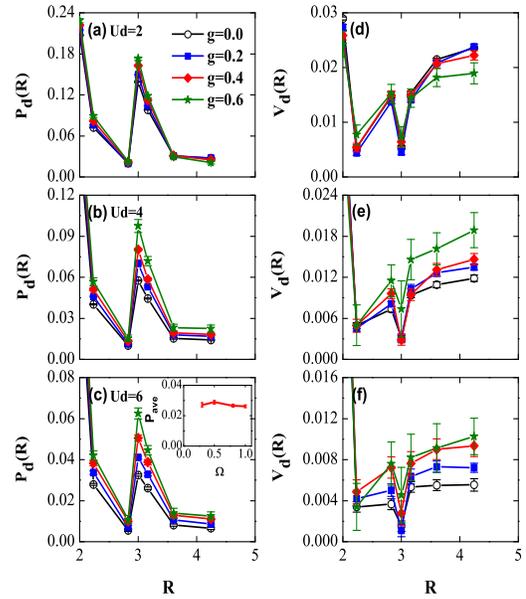,height=9cm,width=8cm,angle=0}
\caption{(color online) $P_{d}(R)$ as a function of the distance $R$
between hole pairs on the $6 \times 6$ cluster at $U_{d}=2$ (a),
$U_{d}=4$ (b) and $U_{d}=6$ (c). (d)-(f) show the vertex
contribution $V_{d}(R)$ corresponding to the left panels. Results
are obtained for $Nh=42$ holes. The el-ph coupling $g$ is indicated
by the shape of the symbol.
The inset of figure (c) displays the average $P_{ave}$ (see text)
as a function of $\Omega$.
} \label{PdVdU246}
\end{figure}
\end{center}

Based on the ED results on the small cluster, we can conclude that
the coupling of AOP to holes can enhance superconductivity on the
$CuO_2$ plane. The question arising is whether the d-wave pairing
symmetry is enhanced? This issue can be addressed by examining the
behavior of the d-wave pairing correlation $P_{d}({\vec R})$ in
Eq.(\ref{pairing}). In Figs.~\ref{PdVdU246}(a)-(c) we show
$P_{d}(R)$ as a function of $R$ for the $6 \times 6$ cluster at
$U_{d}=2$, $4$ and $6$, respectively. The corresponding vertex
contribution is also displayed in Figs.~\ref{PdVdU246}(d)-(f). In
the weak-correlation case ($U_{d}=2$), we observe that $P_{d}(R)$ is
modified slightly by the el-ph coupling. However, when $U_{d}$ is
increased to $4$ and $6$, the pairing correlation is enhanced at all
long-range distances for $R > 2$. At $g=0.6$, the average
enhancement of the long-range part of $P_d$ is estimated to be about
$52\%$ and $65\%$ for $U_{d}=4$ and $U_{d}=6$, respectively. This
increasing enhancement of $P_{d}$ with $U_d$ is consistent with our
findings for the binding energy and quasiparticle weight. A dramatic
increment of the vertex contribution with increasing $g$, as shown
in Fig.~\ref{PdVdU246}(e) and Fig.~\ref{PdVdU246}(f), further
provides strong evidence that the d-wave pairing interaction is
actually increased by the el-ph coupling.

We also study the effect of the tunneling frequency $\Omega$ on the
pairing correlations. In the inset of Fig.~\ref{PdVdU246} (c), the
average of long-range d-wave pairing correlation,
$P_{ave}=(1/N')\sum_{R>2} P_{d}({\vec R})$ where $N'$ is the number
of hole pairs with $R>2$, is plotted as a function of $\Omega$ for
$U_{d}=6$ and $g=0.4$. We notice that its frequency dependence is
rather weak, suggesting that the isotope effect of apical oxygen on
the superconductivity is very small. This is in good agreement with
the site-selected oxygen isotope effect in
$YBa_{2}Cu_{3}O_{6+x}$~\cite{zech}.

\begin{center}
\begin{figure}
\epsfig{file=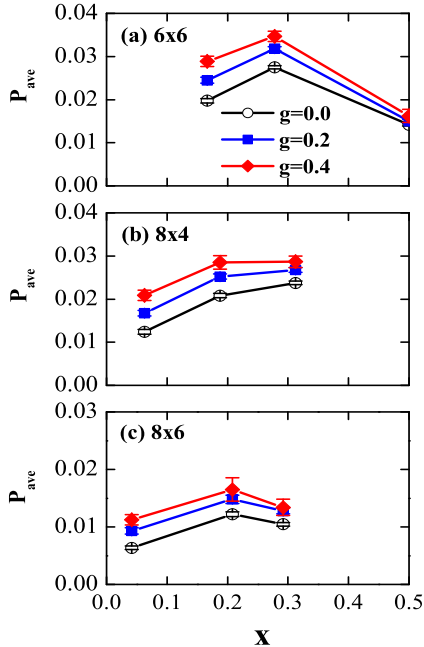,height=10cm,width=7.5cm,angle=0}
\caption{(color online) $P_{ave}$ vs the hole doping density
$x$ for the $6 \times 6$ (a),
$8 \times 4$ (b), and $8 \times 6$ (c) clusters at $U_{d}=6$.}
\label{Pdave}
\end{figure}
\end{center}

To illustrate the effect of hole doping on the phonon-induced
enhancement of dSC, in Fig.~\ref{Pdave} we show $P_{ave}$ as a
function of hole doping density $x$ at $U_{d}=6$ for the $6 \times
6$, $8 \times 4$, and $8 \times 6$ clusters. The combination of
the results on the three clusters shows that in a wide hole doping
region $0.0 < x < 0.5$, the d-wave pairing correlation is
enhanced by the el-ph coupling, and as the hole doping density is
increased from underdoping to optimal doping and then to overdoping,
the enhancement of $P_{ave}$ is weakened monotonically. The reduced
enhancement of $P_{ave}$ with larger hole doping, together with the
stronger enhancement of $\Delta$ and $P_d$ with increasing $U_d$,
demonstrates that strong electronic correlations and/or AFM
fluctuations play a crucial role in the phonon-induced enhancement
of dSC.

\section{\label{Conclusions} Conclusions}
In summary, our numerical simulations show that the hole binding
energy is strongly enhanced by an AOP-induced reduction of
electronic potential energy and phonon related energy. As a
combination of two concurring effects, i.e. the enhancement of hole
pairing interaction and the increase of quasiparticle weight, the
long-range part of d-wave pairing correlations is dramatically
enhanced with increasing the el-ph coupling strength.
Our results also show that strong electronic correlations and/or
AFM fluctuations are crucial for this phonon-induced
enhancement effect. The consistent behavior of our results on
different clusters suggests that the phonon-induced enhancement of
dSC could survive in the thermodynamic limit.

We thank W. Hanke, F.~F. Assaad., and A.~S. Mishchenko for
enlightening discussions. This work was supported by NSFC under
Grant Nos. 10674043 and 10974047. HQL acknowledges support from
HKRGC 402109. EA was supported by the Austrian Science Fund (FWF)
under grant P18551-N16.

\end{document}